\documentclass[aps,prl,preprint,amsmath,amssymb,showpacs,superscriptaddress,nofootinbib]{revtex4-1}
\usepackage{CJKutf8}
\usepackage{graphicx}
\usepackage{dcolumn}
\usepackage{bm}
\usepackage{color}
\usepackage{float}
\usepackage{hyperref}

\allowdisplaybreaks[4]

\begin{document}


\title{Emergence of High-Order Deformation in Rotating Transfermium Nuclei: A Microscopic Understanding}

\author{F. F. Xu}
\affiliation{State Key Laboratory of Nuclear Physics and Technology, School of Physics, Peking University, Beijing 100871, China}
\author{Y. K. Wang}
\affiliation{State Key Laboratory of Nuclear Physics and Technology, School of Physics, Peking University, Beijing 100871, China}
\author{Y. P. Wang}
\affiliation{State Key Laboratory of Nuclear Physics and Technology, School of Physics, Peking University, Beijing 100871, China}
\author{P. Ring}
\affiliation{Physik-Department der Technischen Universit\"at M\"unchen, D-85748 Garching, Germany}
\affiliation{State Key Laboratory of Nuclear Physics and Technology, School of Physics, Peking University, Beijing 100871, China}
\author{P. W. Zhao}
\email{pwzhao@pku.edu.cn}
\affiliation{State Key Laboratory of Nuclear Physics and Technology, School of Physics, Peking University, Beijing 100871, China}

\date{\today}
\begin{abstract}
The rotational properties of the transfermium nuclei are investigated in the full deformation space by implementing a shell-model-like approach in the cranking covariant density functional theory on a three-dimensional lattice, where the pairing correlations, deformations, and moments of inertia are treated in a microscopic and self-consistent way.
The kinematic and dynamic moments of inertia of the rotational bands observed in the transfermium nuclei $^{252}$No, $^{254}$No, $^{254}$Rf, and $^{256}$Rf are well reproduced without any adjustable parameters using a well-determined universal density functional.
It is found for the first time that the emergence of the octupole deformation should be responsible for the significantly different rotational behavior observed in $^{252}$No and $^{254}$No.
The present results provide a microscopic solution to the long-standing puzzle on the rotational behavior in No isotopes, and highlight the risk of investigating only   the hexacontetrapole ($\beta_{60}$) deformation effects in rotating transfermium nuclei without considering the octupole deformation.
\end{abstract}
\maketitle

\date{today}

The synthesis of superheavy nuclei (SHN) toward the predicted ``island of stability'' is in the focus of current nuclear physics research~\cite{Hofmann2000RMP,Oganessian2007JPG}.
The single-particle structure of SHN plays a crucial role in determining the location of the ``island of stability.''
In recent years, the SHN with atomic number from $Z = 113$ to 118 have been discovered~\cite{Munzenberg2015NPA,Oganessian2015NPA,Morita2015NPA}.
However, owing to the very small production cross sections for SHN, so far, only the basic properties such as the dominant decay modes and the lifetimes are known, while the single-particle structure of SHN could not be directly obtained in experiment.
Fortunately, owing to deformation effects, the single-particle orbitals which determine the location of the ``island of stability'' come close to the Fermi surface in the transfermium nuclei located near the deformed shell gaps at $Z=100$ and $N=152$~\cite{Chasman1977RMP}.
Therefore, the spectroscopic experiments on the transfermium nuclei can provide important information for the single-particle structure of SHN.

In-beam, isomer, and decay spectroscopic experiments have been carried out to study the rotational bands in the transfermium nuclei~\cite{Ackermann2017PS}, such as $^{253}$Fm~\cite{Asai2005PRL}, $^{251}$Md~\cite{Chatillon2007PRL}, $^{252-254}$No~\cite{Herzberg2001PRC,Reiter2005PRL,Reiter1999PRL,Butler2002PRL,Tandel2006PRL},
$^{255}$Lr~\cite{Ketelhut2009PRL}, and $^{254,256}$Rf~\cite{Seweryniak2023PRC,Greenlees2012PRL}.
Especially, the rotational bands of the two neighboring nuclei $^{252}$No ($N=150$) and $^{254}$No ($N=152$) have attracted great attention, because they exhibit a  significantly different rotational behavior at high angular momenta~\cite{Herzberg2001PRC}.
The underlying mechanism, however, has been a long-standing puzzle.
Investigations using the total Routhian surface (TRS) method~\cite{Liu2012PRC} and the particle-number-conserving cranked shell model (PNC-CSM)~\cite{Zhang2013PRC} speculate that the different rotational behavior might originate from the hexacontetrapole deformation $\beta_{60}$.
However, both methods neglect the reflection-asymmetric deformation and include only the $\beta_{20}$, $\beta_{40}$, and $\beta_{60}$ degrees of freedom, and they need to fit the data in one way or another.
In addition, the rotational properties of $^{252}$No and $^{254}$No have also been investigated by the macroscopic-microscopic model~\cite{Sobiczewski2001PRC,Zhang2012PRC}, the projected shell model~\cite{Sun2008PRC,Khudair2009PRC}, the reflection asymmetric shell model~\cite{Chen2008PRC}, the nonrelativistic density functional theories~\cite{Duguet2001NPA,Bender2003NPA,Delaroche2006NPA}, and the covariant density functional theories~\cite{Afanasjev2003PRC,Afanasjev2014PS}, but none of these studies have explained the underlying mechanism of the different rotational behavior in $^{252}$No and $^{254}$No.
As a result, it has still been a challenging puzzle up to now.

In this Letter, this long-standing puzzle is microscopically solved for the first time by using the cranking covariant density functional theory (CDFT) in three-dimensional (3D) lattice space with a shell-model-like approach (SLAP).
The cranking CDFT in 3D lattice space~\cite{Ren2019SCI,Ren2020NPA,Zhang2022PRC} provides a microscopic and self-consistent method to calculate the rotational spectra of nuclei in the full deformation space.
Moreover, the SLAP~\cite{Zeng1983NPA,Meng2006FPC,Shi2018PRC,Wang2023PLB,Xu2023IJMPE} is implemented to take into account the pairing correlations with exact particle number conservation, which is crucial for describing the rotational properties of nuclei.
Combining the merits of both CDFT and SLAP, the method is used to study the rotational spectra of $^{252}$No and $^{254}$No as well as the neighboring $^{254}$Rf ($N=150$) and $^{256}$Rf ($N=152$).
The kinematic and dynamic moments of inertia (MOIs) for all nuclei are well reproduced without any adjustable parameters.
A reliable explanation is provided for the puzzling rotational behavior in $^{252}$No and $^{254}$No: the emergence of the octupole deformation should be responsible for the significantly different MOIs at high angular momenta.
Our results highlight the importance of the octupole deformation for describing the rotating transfermium nuclei.
The risk of studying only the $\beta_{60}$ deformation without considering the octupole deformation in rotating transfermium nuclei is also clearly addressed.

The starting point of the CDFT is a universal nuclear energy density functional~\cite{Ring1996PPNP,Vretenar2005PR,Meng2006PPNP,Niksic2011PPNP}.
In the cranking CDFT, the energy density functional is transformed into a frame rotating with a constant rotational velocity $\bm\omega$ around the rotational axis.
The corresponding Kohn-Sham equation for the nucleons is a Dirac equation,
\begin{eqnarray}
  \label{eq:1}
  \hat{h}_0\psi_{k}=\left\{\bm{\alpha}\cdot\left[-i\bm{\nabla}-\bm{V}(\bm{r})\right] + \beta\left[m+S(\bm{r})\right]+V(\bm {r})-\bm\omega\cdot\bm{\hat{J}}\right\}\psi_{k}=\varepsilon_{k}\psi_{k},
\end{eqnarray}
where the cranking single-particle states are $\psi_{k}(\bm r)=\langle\bm r|\hat{b}_{k}^{\dagger}|0\rangle$, and $\varepsilon_{k}$ is the corresponding single-particle Routhian.
The scalar $S(\bm{r})$ and vector $V^\mu(\bm{r})$ fields are connected in a self-consistent way to the nucleon densities and currents; for details see Refs.~\cite{Zhao2011PRL,Zhao2012PRC,Meng2013FP,Zhao2018IJMPE}.
The Dirac equation~\eqref{eq:1} is solved in a 3D lattice space~\cite{Li2020PRC,Xu2024PRC}.

The pairing correlations are taken into account by the SLAP~\cite{Xu2023IJMPE}.
The cranking many-body Hamiltonian with pairing correlations reads as
\begin{eqnarray}
  \label{eq:2} \hat{H}=\hat{H}_0+\hat{H}_{{\rm pair}}.
\end{eqnarray}
The one-body part $\hat{H}_0$ reads as
\begin{eqnarray}
  \label{eq:3} \hat{H}_0=\sum_{m,n>0}\left[h_{mn}\hat{c}_{m}^{\dagger}\hat{c}_{n}+h_{\bar{m}\bar{n}}\hat{c}_{\bar{m}}^{\dagger}\hat{c}_{\bar{n}}\right].
\end{eqnarray}
Here, $\hat{c}_m^\dagger$ represents a set of noncranking single-particle basis obtained from Eq.~\eqref{eq:1} at $\hbar\omega=0$ MeV, and $\bar{m}$ labels the time-reversal state of $m$.
The pairing Hamiltonian $\hat{H}_{{\rm pair}}$ is used for the two-body part,
\begin{eqnarray}
  \label{eq:4}
  \hat{H}_{{\rm pair}}=-G\sum_{m,n>0}\hat{c}_{m}^{\dagger}\hat{c}_{\bar{m}}^{\dagger}\hat{c}_{\bar{n}}\hat{c}_{n},
\end{eqnarray}
where $G$ is the effective pairing strength.
With the transformation coefficients $D_{mk}=\int {\rm d}^3r\langle\bm r|\hat{c}_{m}^{\dagger}|0\rangle^{\ast}\, \langle\bm r|\hat{b}_{k}^{\dagger}|0\rangle$ between the cranking and noncranking single-particle bases, $\hat{H}_{{\rm pair}}$ can be rewritten in the cranking single-particle basis as
\begin{eqnarray}
  \label{eq:5}
    \hat{H}_{{\rm pair}}=-G\sum_{k_1k_2k_3k_4}\sum_{m,n>0}D_{m k_1}^{\ast}D_{\bar{m} k_2}^{\ast}D_{\bar{n} k_4}D_{n k_3}
    \hat{b}_{k_1}^{\dagger}\hat{b}_{k_2}^{\dagger}\hat{b}_{k_4}^{}\hat{b}_{k_3}^{}.
\end{eqnarray}

The cranking many-body Hamiltonian \eqref{eq:2} is diagonalized in the space of many-particle configurations (MPC) $|i\rangle$, which is constructed for the $A$-particle system as
\begin{eqnarray}
  \label{eq:6}
    |i\rangle\equiv|k_1k_2\cdots k_{A}\rangle=\hat{b}_{k_1}^{\dagger}\hat{b}_{k_2}^{\dagger}\cdots\hat{b}_{k_{A}}^{\dagger}|0\rangle,
\end{eqnarray}
and the obtained lowest eigenstate reads as
\begin{eqnarray}
  \label{eq:7}
    \Psi=\sum_{i}C_{i}|i\rangle.
\end{eqnarray}
Here, the coefficients $C_i$ are used to determine the occupation probability for each single-particle state $\psi_{k}$.
Then, the occupation probability for each orbital is iterated back to calculate the densities and currents in the CDFT to achieve self-consistency \cite{Meng2006FPC,Shi2018PRC}.

Finally, one can calculate the pairing energy $E_{{\rm pair}}=\langle\Psi|\hat{H}_{{\rm pair}}|\Psi\rangle$, which is added to the total energy obtained with the CDFT.
The expectation values of the kinematic MOI $J^{(1)}$, the dynamic MOI $J^{(2)}$, and the deformation parameter $\beta_{\lambda\mu}$ can be respectively calculated with
\begin{subequations}
  \begin{eqnarray}
    \label{Eq.8a}J^{(1)}\ &&=\dfrac{\langle\Psi|\hat{J}|\Psi\rangle}{\omega},\\
    \label{Eq.8b}J^{(2)}\ &&=\dfrac{{\rm d}\langle\Psi|\hat{J}|\Psi\rangle}{{\rm d}\omega},\\
    \label{Eq.8c}\beta_{\lambda\mu}\ &&=\dfrac{4\pi}{3AR^{\lambda}}\int d^3\bm{r}\rho_{V}(\bm{r})r^{\lambda}Y_{\lambda\mu}(\Omega),
  \end{eqnarray}
\end{subequations}
where $Y_{\lambda\mu}(\Omega)$ is the spherical harmonics and $\rho_{V}$ is the vector density~\cite{Zhao2012PRC,Meng2013FP,Zhao2018IJMPE}.

In this work, the point-coupling density functional PC-PK1~\cite{Zhao2010PRC} is used.
For the 3D lattice space, the mesh sizes and the grid numbers along the $x$, $y$, and $z$ axes are chosen as 1 fm and 26, respectively.
The dimensions of the cranked MPC space are 1000 for both neutron and proton, and the corresponding neutron and proton pairing strengths are respectively 0.25 MeV and 0.30 MeV.
This provides a nice description of the experimental three-point odd-even mass differences~\cite{Bender2000EPJA}.
A larger MPC space with the renormalized pairing strengths gives essentially the same energy, indicating the convergence of the MPC space.

\begin{table}[!htbp]
  \centering
  \caption{The calculated quadrupole deformations $\beta_{20}$ and $\beta_{22}$, octupole deformations $\beta_{30}$ and $\beta_{32}$, hexadecapole deformation $\beta_{40}$,  hexacontetrapole deformation $\beta_{60}$, and the binding energies $E_{{\rm cal}}$ for the ground states of $^{252}$No, $^{254}$No, $^{254}$Rf, and $^{256}$Rf. The experimental ground-state energies $E_{{\rm exp}}$~\cite{Wang2021CPC} are also listed for comparison. All energies are in MeV.}\label{Tab1}
  \begin{ruledtabular}
  \begin{tabular}{ccccccccc}
    Nucleus & $\beta_{20}$ & $\beta_{22}$ & $\beta_{30}$ & $\beta_{32}$ & $\beta_{40}$ & $\beta_{60}$ & $E_{{\rm cal}}$ & $E_{{\rm exp}}$ \\\hline\rule{0pt}{14pt}
    $^{252}$No & 0.296 & 0.011 & 0.0042 & 0.0128 & 0.082 & $-$0.029 & 1871.3 & 1871.3 \\\rule{0pt}{14pt}
    $^{254}$No & 0.297 & 0.013 & 0.0004 & 0.0000 & 0.062 & $-$0.049 & 1885.6 & 1885.6 \\\rule{0pt}{14pt}
    $^{254}$Rf & 0.300 & 0.000 & 0.0000 & 0.0006 & 0.069 & $-$0.044 & 1876.7 & 1875.5 \\\rule{0pt}{14pt}
    $^{256}$Rf & 0.302 & 0.000 & 0.0004 & 0.0000 & 0.049 & $-$0.063 & 1892.0 & 1890.7 \\
  \end{tabular}
  \end{ruledtabular}
\end{table}

In Table~\ref{Tab1}, the calculated deformation parameters and binding energies for the ground states of $^{252}$No, $^{254}$No, $^{254}$Rf, and $^{256}$Rf are listed in comparison with the experimental data~\cite{Wang2021CPC}.
The experimental binding energies are nicely reproduced within 1.30 MeV.
The $\beta_{20}$ deformations are close to 0.30, which are consistent with the data available~\cite{Herzberg2001PRC,Reiter1999PRL}, while the $\beta_{22}$ deformations are negligible.
In addition to the sizable quadrupole deformations, the $\beta_{40}$ and $\beta_{60}$ deformations are also visible.
It is particularly notable that the nonzero $\beta_{32}$ deformation in $^{252}$No is supported by the appearance of the low-lying $2^{-}$ band in the experimental spectrum~\cite{Sulignano2007EPJA}.

\begin{figure*}[!htbp]
  \centering
  \includegraphics[width=0.9\textwidth]{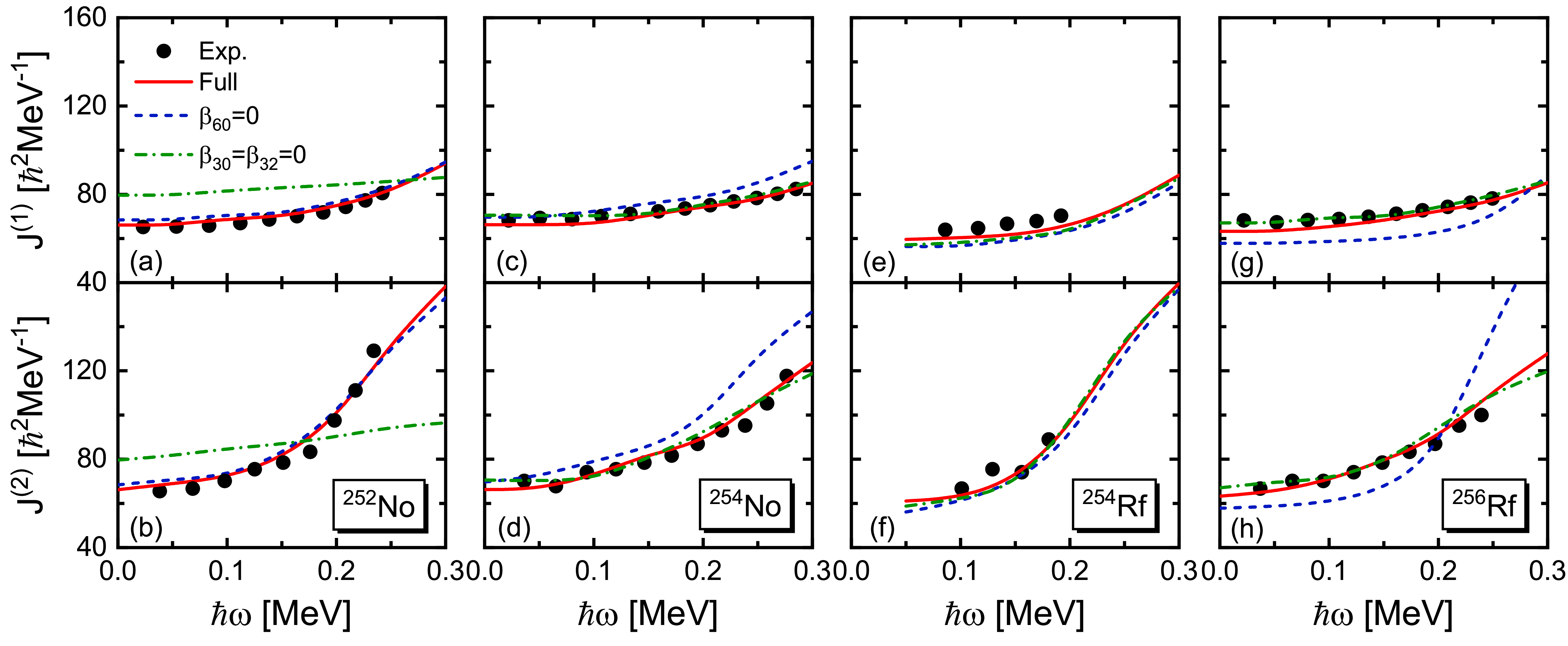}
  \caption{Calculated kinematic MOIs $J^{(1)}$ and dynamic MOIs $J^{(2)}$ for $^{252}$No (a),(b); $^{254}$No (c),(d); $^{254}$Rf (e),(f); and $^{256}$Rf (g),(h) in comparison with the experimental data (symbols)~\cite{Herzberg2001PRC,Reiter1999PRL,Seweryniak2023PRC,Greenlees2012PRL}.
  The results obtained in the full deformation space (Full), imposing only vanishing $\beta_{60}$ deformation ($\beta_{60}=0$), and imposing only vanishing octupole deformations ($\beta_{30}=\beta_{32}=0$) are represented by solid, dashed, and dashed-dotted lines, respectively.
  }
  \label{Fig1}
\end{figure*}

In Fig.~\ref{Fig1}, the calculated kinematic MOIs $J^{(1)}$ and dynamic ones $J^{(2)}$ as functions of the rotational frequency are depicted in comparison with the data~\cite{Herzberg2001PRC,Reiter1999PRL,Seweryniak2023PRC,Greenlees2012PRL}.
Both the kinematic and dynamic MOIs in the full deformation space (red solid lines) agree well with the experimental data.
Note that the experimental dynamic MOIs of $^{252}$No exhibit an abrupt upbending at $\hbar\omega\approx0.15$ MeV, while those of $^{254}$No are much smoother.
The distinct behavior for these two neighboring nuclei is also very well reproduced in the present calculations.

The deformation $\beta_{60}$ plays different roles in reproducing the MOIs in $N=150$ and $N=152$ nuclei.
If the deformation $\beta_{60}$ is constrained to zero (blue dashed lines), the calculated MOIs for $^{252}$No are almost unchanged, and the data are still  reproduced.
For $^{254}$No, however, the calculated MOIs become larger especially for the rotational frequency $\hbar\omega$ higher than 0.15 MeV, where a rapid increase of the dynamic MOIs can be seen.
Qualitatively, this is similar to the results given by the TRS~\cite{Liu2012PRC} and PNC-CSM methods~\cite{Zhang2013PRC}, in which only the deformations $\beta_{\lambda0}$ with even $\lambda$ are considered.
The $\beta_{60}$ deformation effects have also been studied for the MOIs of $^{254}$Rf ($N=150$) and $^{256}$Rf ($N=152$).
Interestingly, it is also found that the $\beta_{60}$ deformation has distinct roles in the Rf isotopes with $N=150$ and $N=152$.

Considering the visible octupole deformations in $^{252}$No, we also study the effects of the octupole deformations on the MOIs for the first time.
If the deformations $\beta_{30}$ and $\beta_{32}$ are constrained to zero (green dashed-dotted lines), the calculated MOIs for $^{252}$No do not agree with the data, indicating the crucial role of the octupole deformations.
However, the octupole deformations have almost no effects on the MOIs for $^{254}$No, $^{254}$Rf, and $^{256}$Rf.
Although the octupole deformation effects on the MOIs are similar in the $N=150$ and $N=152$ Rf isotopes, they are quite different in the corresponding No isotopes.
This suggests that the observed different rotational behavior in the No isotopes may originate from the octupole deformations.

\begin{figure}[!htbp]
  \centering
  \includegraphics[width=0.45\textwidth]{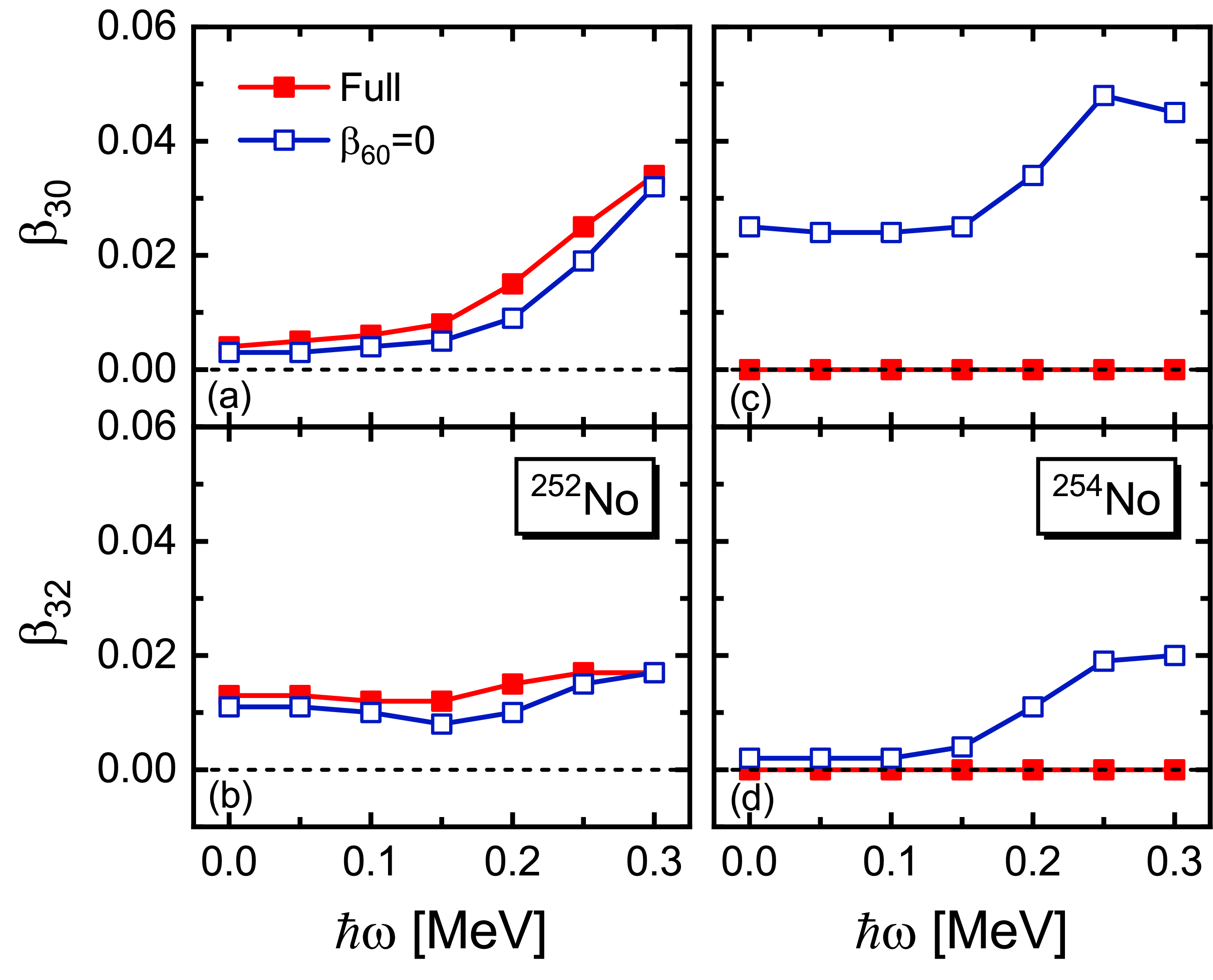}
  \caption{Calculated octupole deformations $\beta_{30}$ and $\beta_{32}$ for $^{252}$No (a),(b) and $^{254}$No (c),(d) with (solid squares) and without (open squares) $\beta_{60}$ deformation.
  }
  \label{Fig2}
\end{figure}

To further explore the octupole deformation effects on the MOIs for $^{252}$No and $^{254}$No, the calculated octupole deformations $\beta_{30}$ and $\beta_{32}$ as functions of the rotational frequency for these two nuclei are shown in Fig.~\ref{Fig2}.
The $\beta_{30}$ and $\beta_{32}$ values (red solid squares) for $^{252}$No keep almost constant at low frequency, and they sharply increase at $\hbar\omega\approx0.15$ MeV, which just corresponds to the upbending point as seen in Figs.~\ref{Fig1}(a)-\ref{Fig1}(b).
In contrast, for $^{254}$No, the calculated $\beta_{30}$ and $\beta_{32}$ values are always zero along the rotational band.
This demonstrates the distinct octupole deformations in $^{252}$No and $^{254}$No.

If the deformation $\beta_{60}$ is fixed to zero (blue open squares), the $\beta_{30}$ and $\beta_{32}$ values for $^{252}$No are nearly unchanged as compared to the ones obtained in the full deformation space. In particular, the surge at $\hbar\omega\approx0.15$ MeV still exists.
However, the octupole deformations for $^{254}$No are dramatically different from those obtained in the full deformation space.
Both $\beta_{30}$ and $\beta_{32}$ deformations become nonzero, and they exhibit a sudden rise at $\hbar\omega\approx0.15$ MeV, which corresponds to the upbending as seen in Figs.~\ref{Fig1}(c)-\ref{Fig1}(d).

Based on these analyses, it is clear that the octupole deformation should be responsible for the upbending of the rotational bands in $^{252}$No.
The octupole deformation of $^{254}$No keeps vanishing along the rotational band, leading to its smooth rotational behavior.
Therefore, the emergence of the octupole deformation in only $^{252}$No explains the distinct rotational behavior observed in these two isotopes.
Moreover, it is indeed risky to investigate only the effects of the deformation $\beta_{60}$ on the rotation behavior without considering the octupole deformations, because the octupole and hexacontetrapole deformations are coupled with each other.

\begin{figure*}[!htbp]
  \centering
  \includegraphics[width=0.9\textwidth]{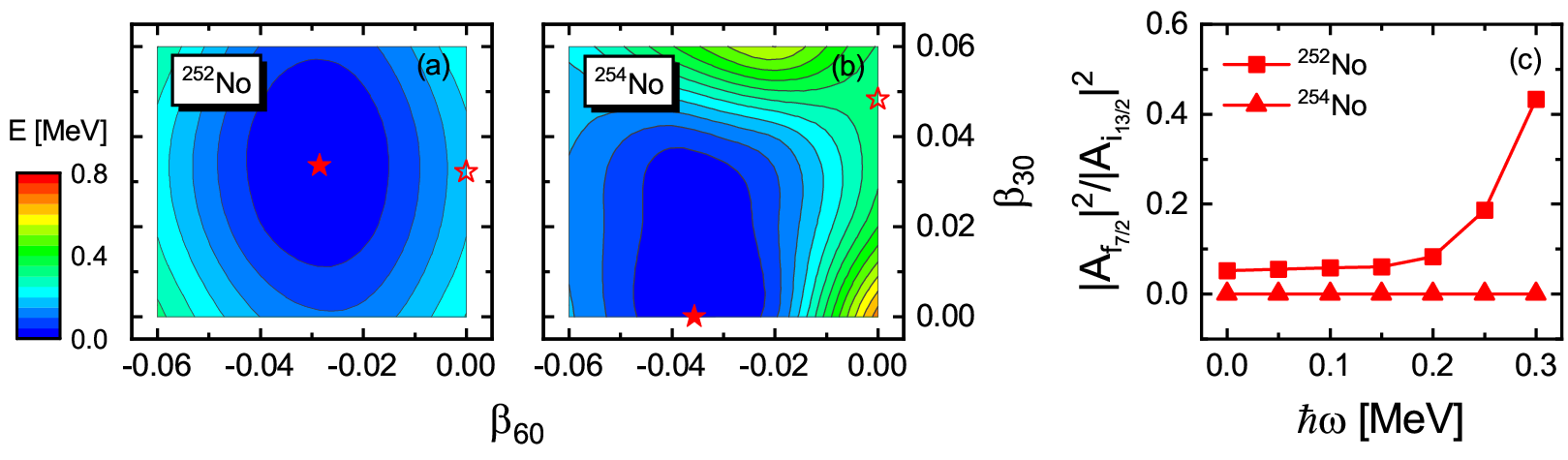}
  \caption{Potential energy surfaces of $^{252}$No (a) and $^{254}$No (b) in the $(\beta_{60},\beta_{30})$ plane at $\hbar\omega=0.3$ MeV. The energy minimum and the ``spurious'' minimum without $\beta_{60}$ deformation are indicated by a solid star and an open star, respectively. The energies are normalized with respect to the minimum energy, and the contour interval is 0.05 MeV.
  (c): The amplitude ratios between the $f_{7/2}$ and $i_{13/2}$ components for the highest-$j$ orbital near the proton Fermi level as a function of the rotational frequency for $^{252}$No (solid squares) and $^{254}$No (solid triangles).
  }
  \label{Fig3}
\end{figure*}

The coupling of the octupole and $\beta_{60}$ deformations can be clearly seen from the potential energy surface in the $(\beta_{60},\beta_{30})$ plane, which is depicted in Figs.~\ref{Fig3}(a)-\ref{Fig3}(b) for the case of $\hbar\omega=0.3$ MeV.
The energy minimum for $^{252}$No has a nonzero $\beta_{30}$ deformation, whereas for $^{254}$No, this minimum has a zero $\beta_{30}$ value and is very soft in the $\beta_{30}$ direction.
Moreover, for $^{254}$No, one can also see for $\beta_{60}=0$ a ``spurious'' minimum with a nonzero $\beta_{30}$ value.
This explains the coupling of the deformation $\beta_{60}$ and the octupole deformations, indicating that it is wrong to investigate the effects of the deformation $\beta_{60}$ on the MOIs without considering the octupole deformations.
It is even more wrong to neglect the octupole deformations for $^{252}$No, because it has a visible $\beta_{30}$ deformation.

To understand the significant difference of the octupole deformations in $^{252}$No and $^{254}$No in a more microscopic way, the components of the single-proton orbitals near the Fermi surface are analyzed by considering the overlap of the wave functions with those of the spherical Woods-Saxon basis~\cite{Koepf1991ZPA}.
The amplitude ratios $|A_{f_{7/2}}|^2/|A_{i_{13/2}}|^2$ between the $f_{7/2}$ and $i_{13/2}$ components for the highest-$j$ proton orbital, which reflect the coupling strength between the proton $i_{13/2}$ and $f_{7/2}$ orbitals, are depicted in Fig.~\ref{Fig3}(c).
For $^{252}$No, the ratios $|A_{f_{7/2}}|^2/|A_{i_{13/2}}|^2$ show a slight increase with the rotational frequency below $\hbar\omega=0.15$ MeV and then a rapid rise, which results in the trend of the octupole deformations as shown in Figs.~\ref{Fig2}(a)-\ref{Fig2}(b).
In contrast, the ratios $|A_{f_{7/2}}|^2/|A_{i_{13/2}}|^2$ for $^{254}$No stay zero for all rotational frequencies, indicating that there is no coupling between the proton $i_{13/2}$ and $f_{7/2}$ orbitals. This explains the vanishing octupole deformations in Figs.~\ref{Fig2}(c)-\ref{Fig2}(d).

In summary, the rotational properties of the transfermium nuclei are investigated in the full deformation space by implementing a shell-model-like approach in the cranking covariant density functional theory on a 3D lattice, where the pairing correlations, deformations, and moments of inertia are treated in a microscopic and self-consistent way.
The kinematic and dynamic moments of inertia of the rotational bands observed in the transfermium nuclei $^{252}$No, $^{254}$No, $^{254}$Rf, and $^{256}$Rf are well reproduced without any adjustable parameters using a well-determined universal density functional.
In particular, the distinct rotational behavior at high angular momenta in $^{252}$No and $^{254}$No is also successfully reproduced.
It is found for the first time that the emergence of the octupole deformation should be responsible for the difference in $^{252}$No and $^{254}$No.
The emergence of the octupole deformation in $^{252}$No is associated with the significant coupling between the proton $i_{13/2}$ and $f_{7/2}$ orbitals, especially at high  angular momenta.
Moreover, the potential energy surface of $^{254}$No in the $(\beta_{60},\beta_{30})$ plane explains the coupling between the hexacontetrapole and octupole deformations, indicating a risk of investigating only the effects of the hexacontetrapole ($\beta_{60}$) deformation in rotating transfermium nuclei without considering the octupole deformations.
The present work provides a microscopic solution to the long-standing puzzle on the rotational behavior in No isotopes, and highlights the importance of the octupole deformation for describing the rotating transfermium nuclei.

\begin{acknowledgments}
 This work was partly supported by  the National Natural Science Foundation of China (Grants No. 11975031, No. 12141501, No. 11935003, No. 12070131001, and No. 12105004), the China Postdoctoral Science Foundation under Grant No. 2020M680183, the High-performance Computing Platform of Peking University, and by the Deutsche Forschungsgemeinschaft (DFG, German Research Foundation) under Germany's Excellence Strategy EXC-2094-390783311, ORIGINS.
\end{acknowledgments}

\end{document}